\documentclass[twocolumn,showpacs,preprintnumbers,amsmath,amssymb,superscriptaddress, prb]{revtex4}
\usepackage{amssymb}
\usepackage{graphicx}
\usepackage{dcolumn}
\usepackage{bm}
\usepackage{times}
\usepackage{hyperref}
\usepackage{color}

\usepackage{textcomp}

\newcommand{\cmto}{Co$_{1-x}$Mg$_x$Ta$_2$O$_6$}
\newcommand{\cto}{CoTa$_2$O$_6$}
\newcommand{\mto}{MgTa$_2$O$_6$}
\newcommand{\xone}{Co$_{0.9}$Mg$_{0.1}$Ta$_2$O$_6$}

\newcommand{\xseven}{Co$_{0.3}$Mg$_{0.7}$Ta$_2$O$_6$}

\newcommand{\muB}{$\mu_\mathrm{B}$}

\begin{document}
%
\title{Signatures of low-dimensional magnetism and short-range magnetic order in Co-based trirutiles}
%
%
%
\author{R. Baral}  
\affiliation{Department of Physics, 500 W University Ave, University of Texas at El Paso, El Paso, TX 79968, USA}
\author{H. S. Fierro}  
\affiliation{Department of Physics, 500 W University Ave, University of Texas at El Paso, El Paso, TX 79968, USA}
\author{C. Rueda}  
\affiliation{Department of Physics, 500 W University Ave, University of Texas at El Paso, El Paso, TX 79968, USA}
\author{B. Sahu}  
\affiliation{Highly Correlated Matter Research Group, Department of Physics, University of Johannesburg, P. O. Box 524, Auckland Park 2006, South Africa}
\author{A. M. Strydom} 
\affiliation{Highly Correlated Matter Research Group, Department of Physics, University of Johannesburg, P. O. Box 524, Auckland Park 2006, South Africa}
\author{N. Poudel}
\affiliation{Idaho National Laboratory, Idaho Falls, ID 83415, USA}
\author{K. Gofryk}
\affiliation{Idaho National Laboratory, Idaho Falls, ID 83415, USA}
\author{F. S. Manciu}  
\affiliation{Department of Physics, 500 W University Ave, University of Texas at El Paso, El Paso, TX 79968, USA}
\author{C. Ritter} 
\affiliation{Institut Laue Langevin, 71, Avenue des Martyrs, Grenoble 38000, France}
\author{T. W. Heitmann}
\affiliation{University of Missouri Research Reactor, University of Missouri, Columbia, MO 65211, USA}
\author{B. P. Belbase$^\dagger$}
\affiliation{Central Department of Physics, Tribhuvan University, Kirtipur, 44613, Kathmandu, Nepal}
\affiliation{Condensed Matter Physics Research Center, Butwal, Rupandehi, Nepal}
\author{S. Bati$^\dagger$}
\affiliation{Central Department of Physics, Tribhuvan University, Kirtipur, 44613, Kathmandu, Nepal}
\affiliation{Condensed Matter Physics Research Center, Butwal, Rupandehi, Nepal}
\author{M. P. Ghimire}
\affiliation{Central Department of Physics, Tribhuvan University, Kirtipur, 44613, Kathmandu, Nepal}
\affiliation{Leibniz Institute for Solid State and Materials Research IFW Dresden, 01069 Dresden, Germany}
\affiliation{Condensed Matter Physics Research Center, Butwal, Rupandehi, Nepal}
\author{H. S. Nair$^*$}
\affiliation{Department of Physics, 500 W University Ave, University of Texas at El Paso, El Paso, TX 79968, USA}
\date{\today} 
%
%
%

\begin{abstract}
Features of low dimensional magnetism resulting from 
a square-net arrangement of Co atoms in trirutile \cto\ 
is studied in the present work by means of density functional theory
and is compared with the experimental results
of specific heat and neutron diffraction.
The small total energy differences between the ferromagnetic (FM)
and antiferromagnetic (AFM) configuration of \cto\ shows 
that competing magnetic ground states exist, with the possibility 
of transition from FM to AFM phase at low temperature. 
Our calculation further suggests the semi-conducting 
behavior for \cto\ with a band gap of $\sim$0.41~eV.
The calculated magnetic anisotropy energy is
$\sim$2.5~meV with its easy axis along the [100] (in-plane) direction.
Studying the evolution of magnetism in \cmto\ ($x$ = 0, 0.1, 0.3, 0.5, 0.7 and 1).\, 
it is found that the sharp AFM transition exhibited by
\cto\ at $T_N$ = 6.2~K in its heat capacity vanishes with Mg-dilution,
indicating the obvious effect of weakening the 
superexchange pathways of Co.
The current specific heat study reveals the robust nature
of $T_N$ for \cto\ in applied magnetic fields.
Clear indication of short-range magnetism is obtained
from the magnetic entropy, however, diffuse components
are absent in neutron diffraction data.
At $T_N$, \cto\ enters a long-range ordered magnetic state
which can be described using a propagation vector, ($\frac{1}{4} \frac{1}{4} 0$).
Upon Mg-dilution at $x\geq$0.1, the long-range
ordered magnetism is destroyed.
The present results should motivate an investigation of
magnetic excitations in this low-dimensional anisotropic magnet.    
\end{abstract}
\maketitle
\section{Introduction}
\cto\ belongs to the class of $MT_2$O$_6$ ($M$ = Ni, Fe, Co; $T$ = Ta, Sb) 
low-dimensional magnetic oxide compounds in trirutile structure.
One of the earliest studies on \cto\ 
reports a broad transition at 14~K in magnetization 
and relates it to low-dimensional physics 
\cite{takano_magnetic_1970}.
Specific heat of \cto\ was studied
early on and was analyzed based on 
the model of an Ising net of spins, again
making a connection to low dimensionality \cite{kremer_specific_1988}.
Subsequent studies by other groups have identified reduced
dimensionality of magnetic interactions present in
trirutiles that adopt the $P4_2/mnm$ space group where the transition metal
is situated inside an octahedral environment produced by oxygen ligands.
The magnetic superexchange pathways are dictated by the
one dimensional $M \textendash O \textendash O \textendash M$ chains
along [110] at $z$ = 0 and [1$\overline 1$0] at $z$ = $\frac{1}{2}$,
which renders low dimensionality to the Co spin system
\cite{reimers_crystal_1989, whangbo2003m}.
The importance of this class of low-dimensional magnets has been
put forward with respect to antiferromagnets with reduced dimensionality
that can realize Luttinger liquids \cite{haldane1981luttinger}
and one dimensional antiferromagnets in external magnetic fields
that can realize spin liquids \cite{dmitriev2004jetp, dmitriev2004prb, kurmann1981j, kurmann1982j}. 
Strongly correlated magnetic behaviour with reduced dimensionality
has been seen in a related trirutile tantalate, NiTa$_2$O$_6$ \cite{law_strongly_2014}.
Experimental and theoretical support for low dimensionality 
and anisotropic exchange interactions parallel and perpendicular 
to spin chains are found in several related systems 
\cite{kato_one-dimensional_2000, christian_magnetic_2014, rebello_transition_2013, kasinathan2008quasi, christian2017local}.
\\
\begin{figure}[!b]
	\includegraphics[scale=0.28]{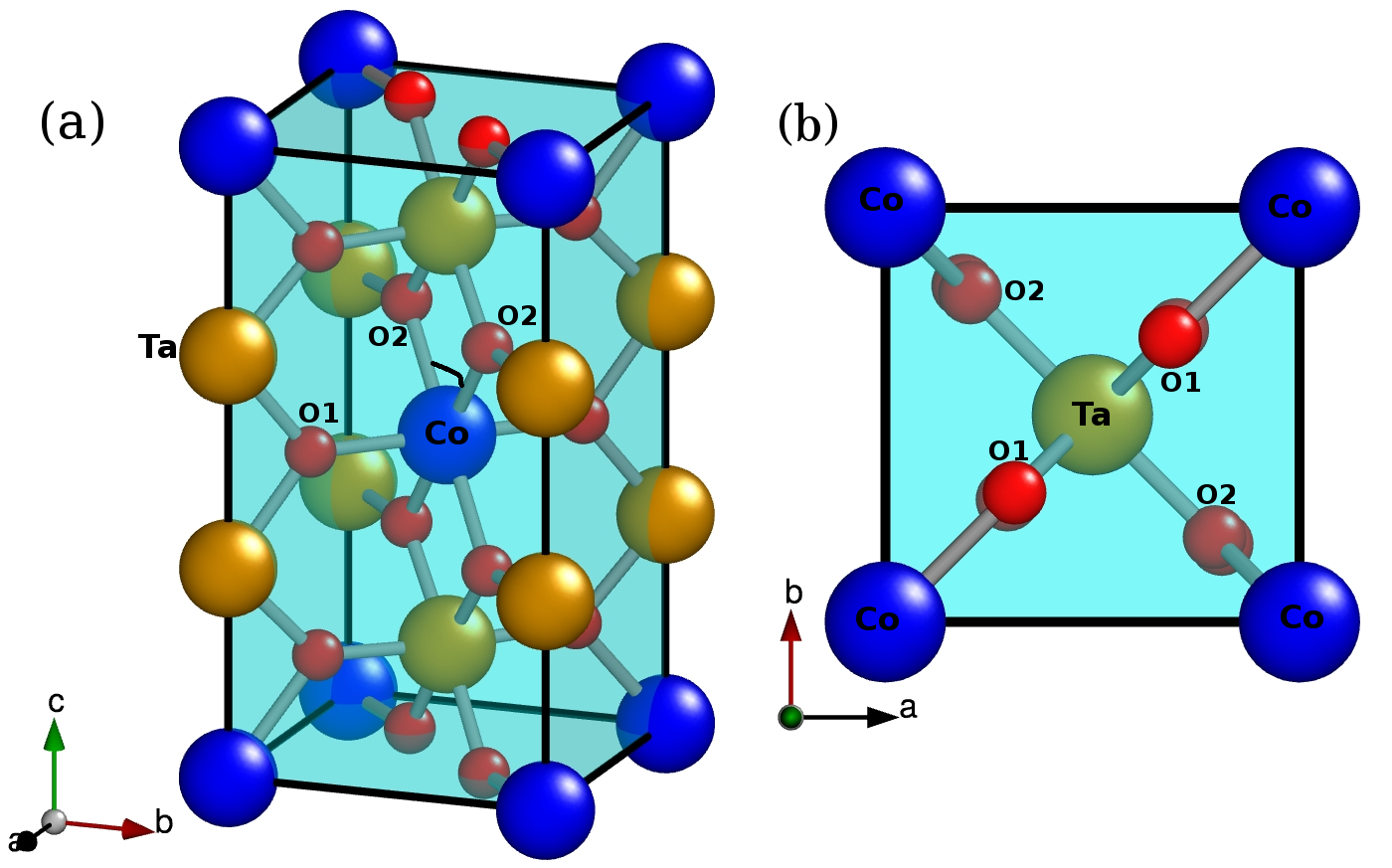}
	\caption{(color online) \label{fig:str} 
		The crystal structure of \cto. 
		(a) The Co atoms represented by blue spheres forms CoO$_6$ octahedra with the
		oxygen atoms (in red). (b) Shows a projection on to the $ab$-plane
		from which the square planar arrangement of Co can be seen. The
		figures were created using xfplo version of FPLO \cite{klaus}.}
\end{figure}
\indent
Recent investigation on \cto\ has revealed optical
dichroism in the single crystals \cite{christian_magnetic_2018}.
The dichroism was attributed to the anisotropic 
exchange mechanism related to the  
$M \textendash O \textendash O \textendash M$ chains.
In addition to the low-dimensionality and anisotropic exchange,
the specific heat of \cto\ presented evidence for a large
fraction of spins of Co remaining disordered even at 
very low temperatures below the $T_N$ (6.1~K) 
\cite{christian_magnetic_2018}.
The value of entropy change, 4.11~J/mol-K, was estimated
to be below the maximum value for $S$ = 1/2.
However, this must be seen against a significant 
amount of entropy release in \cto\
occurring above $T_N$.
Effects of anisotropy are reported in the specific heat of
\cto\ measured with external magnetic field applied
parallel and perpendicular to the spin chains of Co.
When the applied field was parallel to [110], a broad peak
emerged below the main peak at $T_N$. 
While the peak at $T_N$ remained robust up to 8~T, 
the broad peak shifted its position under the application of field. 
On the other hand, when the field was parallel to [100], 
only one peak was detected in the specific heat.
The anisotropy in magnetization and specific heat
contributed towards a moderate magnetocaloric effect in \cto 
\cite{christian_magnetic_2018}.
An anisotropic magnetocaloric effect has been reported 
in NiTa$_2$O$_6$ \cite{christian_magnetic_2014}.
Spin-phonon interactions and their signatures in low-dimensional
magnetic systems were recently probed in the case of $MT_2$O$_6$
($M$ = Ni, Co; $T$ = Ta, Sb) \cite{prasai2018resonant}.
Significant reduction in the thermal conductivity of the Ta-based
compounds was observed compared to that of the Sb-based systems.
The thermal conductivity in these compounds was described
based on resonant absorption of phonons in a two-level system.
In order to clearly understand the connection between the spin and
lattice entities, an unambiguous determination of the crystal
and magnetic structures is a prerequisite.
\\
\indent 
Existing reports on the magnetic structure of \cto\
describe slightly differing structures 
\cite{kinast_bicriticality_2003, kinast_magnetic_2010, reimers_crystal_1989}.
The most recent estimation of magnetic structure
is that of an antiferromagnetic structure with the propagation vector 
($\pm \frac{1}{4}$ $\frac{1}{4}$ $\frac{1}{4}$)
where the magnetic moments lie entirely on the Co$\textendash$O planes
leading to a two dimensional magnetic structure \cite{kinast_magnetic_2010}.
In an earlier study of the solid solution Fe$_x$Co$_{1-x}$Ta$_2$O$_6$
using neutron diffraction, a double-$\mathrm k$ magnetic structure
with  ($\pm \frac{1}{4}$ $\frac{1}{4}$ $\frac{1}{4}$) was reported for
\cto\ while FeTa$_2$O$_6$ had  ($\frac{1}{2}$ 0 $\frac{1}{2}$) and 
(0 $\frac{1}{2}$ $\frac{1}{2}$) \cite{kinast_bicriticality_2003}.
This study pointed out the existence of a bicritical point in the 
$xT$ phase diagram of the mixed composition.
One of the earliest neutron diffraction studies of \cto\ had ascribed
the observed magnetic reflections to a propagation vector
($\frac{1}{4}$ $\frac{1}{4}$ $\frac{1}{4}$) 
and used a two-cone axis helical spin structure to
explain its magnetic structure \cite{reimers_crystal_1989}.
In the present report, we extend our preliminary investigation using
magnetization \cite{baral2019low} to include a detailed study of the
\cmto\ compounds using neutron diffraction, specific heat, 
Raman scattering, thermal conductivity and density functional 
theory computations.
Our work provides support to the previous findings on low-dimensional
magnetism but also improves the determination of the magnetic structure
through neutron diffraction. 
The importance of inherent short-range spin dynamics is highlighted
to warrant future neutron spectroscopic work.
\section{Experimental methods}
\indent
The polycrystalline powder samples used in the present study
were prepared following the prescriptions used in
a previous work \cite{baral2019low}.
Neutron diffraction experiments were carried out 
on 4~g powders, at the PSD instrument in University of Missouri 
Research Reactor (MURR) using $\lambda$ = 1.485~{\AA}.
The sample powders were loaded in a vanadium can which was placed 
in an Al holder purged and filled with helium. 
Diffraction patterns were collected at different temperatures
including 295~K, 7~K and 5~K for different compositions.
Fullprof suite of programs\cite{fullprof} was used to
perform the Rietveld analysis\cite{rietveld} of the
diffractograms. 
Determination of the magnetic structure was performed 
using representation analysis by way of the software 
SARA$h$ \cite{wills_sarah}.
The specific heat, $C_p(T)$, of the samples (2-3~mg pellets) 
in the range 2~K$\textendash$300~K under 0~T and 7~T
was measured using the heat pulse method in a commercial
Physical Property Measurement System from Quantum Design.
The thermal conductivity was measured by pulse-power method 
using the Thermal Transport Option (TTO) and a 
Quantum Design DynaCool-9 system.
The Raman scattering data were acquired at ambient conditions in 
backscattering geometry with an alpha 300R WITec system 
(WITec GmbH, Ulm, Germany). 
A 532~nm excitation of a frequency-doubled neodymium-doped 
yttrium–aluminum–garnet (Nd:YAG) 
laser that was restricted to a power output of a 
few mW, and a 20 X objective lens with a numerical 
aperture of 0.4 were used. 
The accumulation of data for each sample consisted of 
20 Raman spectra, with each spectrum acquired in 
500 milliseconds, for a total time 
acquisition of 10 seconds. 
The spectral resolution was 4~cm$^{-1}$. 
Appropriate background subtractions were 
performed for all Raman spectra.
\section{Computational Details}
Density functional theory (DFT) calculations were performed 
to study the electronic and magnetic properties of 
\cto with the full-potential local-orbital (FPLO) code 
\cite{klaus}, version 18.00. 
The exchange and correlation energy considered is the 
generalized gradient approximation (GGA) in the 
parametrization of Perdew, Burke, and Ernzerhof 
(PBE-96) \cite{pbe}. 
Self-consistent calcualtions were carried out using 
the scalar and four-component full relativistic 
mode of FPLO.
The basis states that were treated as valence states are: 
Co: 3$s$, 3$p$, 4$s$, 5$s$, 3$d$, 4$d$, 4$p$ and Ta: 4$s$, 4$p$, 4$d$, 5$s$, 
6$s$, 5$d$, 5$p$, 6$p$, and O: 2$s$, 2$p$, 3$d$. 
The $k$-space integrations were carried out with the 
linear tetrahedron method using a 12$\times$12$\times$12 
$k$-mesh in the full Brillouin zone. 
The convergence criteria for self consistency 
cycle were set to be 10$^{-8}$ H for energy convergence 
and 10$^{-6}$ electronic charge for the charge convergence. 
The calculations obtained from FPLO were re-checked using 
the full-potential linearized augmented plane wave (FP-LAPW) 
method as implemented in the WIEN2k code \cite{blaha}. 
The choice of the muffin-tin radii of Co, Ta, and 
O atoms were taken as 2.07, 2.0 and 1.72  Bohr respectively. 
\section{Results and Discussion}
\subsection{Specific heat}
\begin{figure}[!b]
	\includegraphics[scale=0.32]{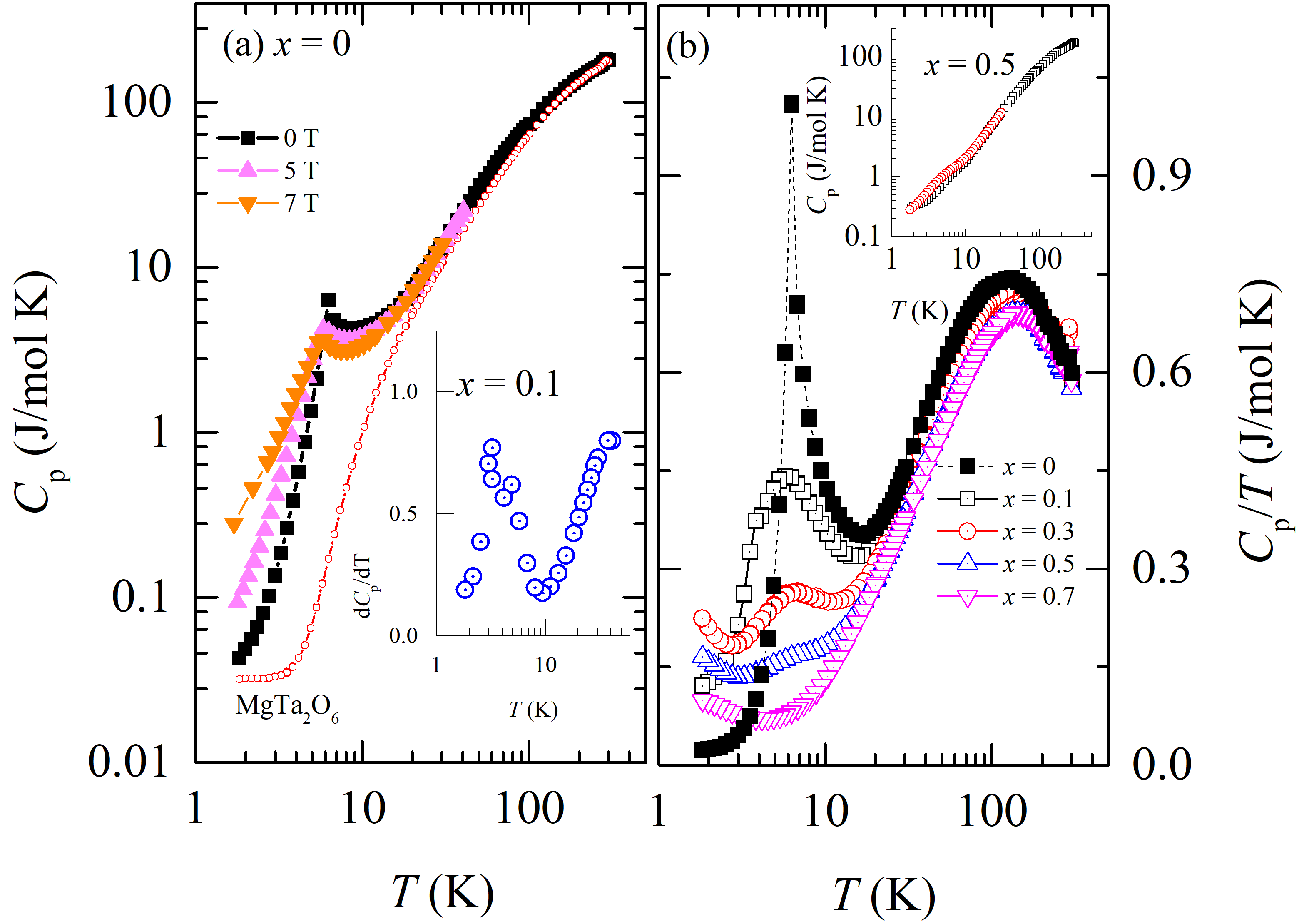}
	\caption{ \label{fig_cp} (color online) 
		(a) The specific heat of \cto\ shows the magnetic
		phase transition at $T_N$ = 6.2~K. 
		The peak at $T_N$ appears robust for 5~T and 7~T. 
		As a lattice analogue, the specific heat of \mto\ is presented in red open circles. 
		$dC_p/dT$ of $x$ = 0.1 is shown in the inset.
		(b) The plot of $C_p/T$ versus $T$ at 0~T for the 
		different compositions of \cmto. 
		The sharp transition in the case of $x$ = 0 vanishes upon increasing $x$. 
		The inset shows the specific heat of $x$ = 0.5 in 0~T and 7~T.}
\end{figure}

\begin{figure}[!t]
	\includegraphics[scale=0.25]{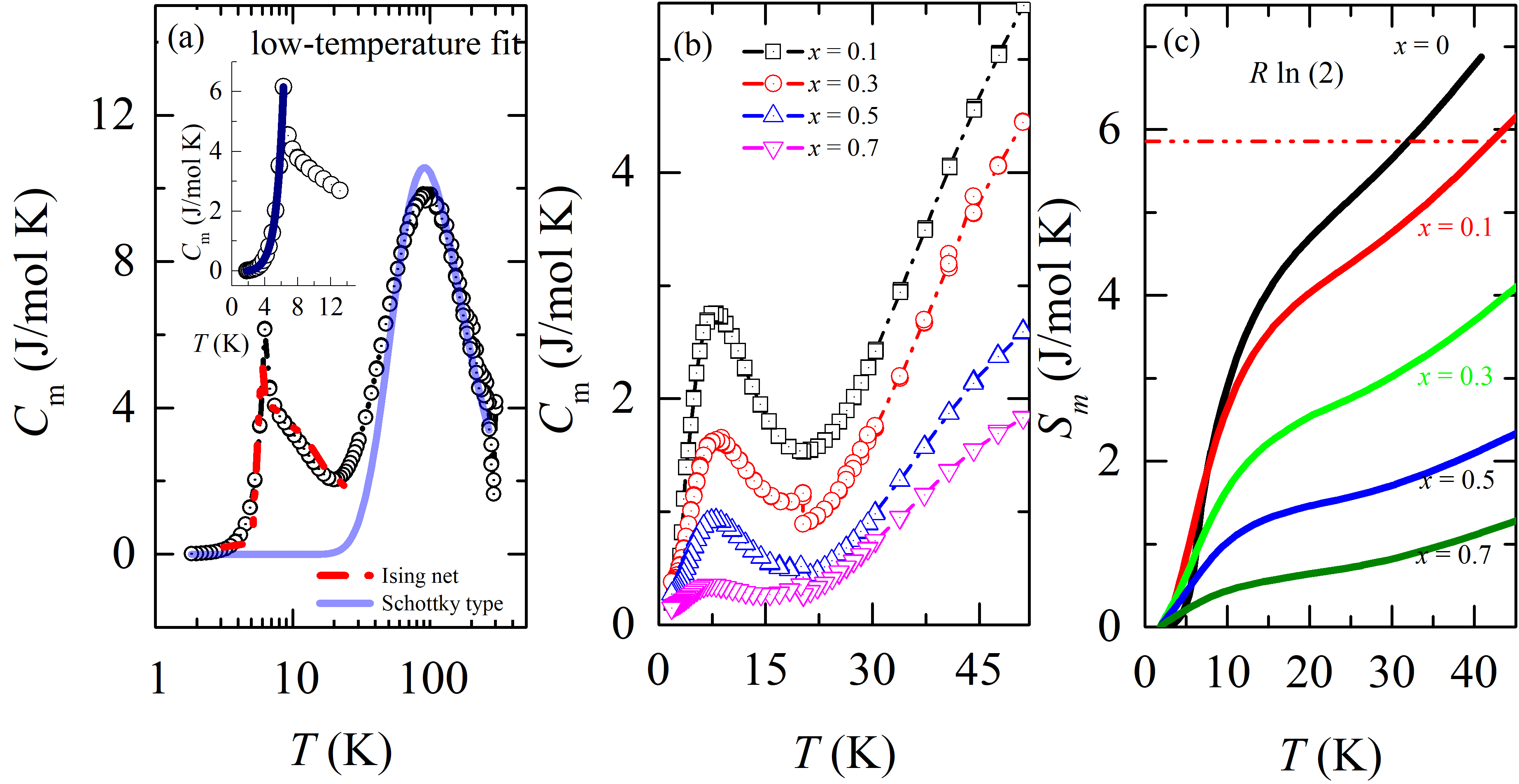}
	\caption{ \label{fig_cp2} (color online) 
		(a) The magnetic specific heat $C_m$ of \cto\ (open circles). 
		The red solid line is the specific heat of a square net of Ising spins
		while the blue line is a fit using a two-level Schottky term.
		The inset shows the $C_m$ along with a fit using Eqn.~(\ref{eqn1}). 
		(b) Shows $C_m$ of $x$ = 0.1, 0.3, 0.5 and 0.7. 
		The magnetic entropy, $S_m$, for \cmto\ $x$ = 0 to 0.7 are plotted in (c).}
\end{figure}

\indent
The specific heat curves, $C_p(T)$, of \cto\ obtained
under the application of 0~T, 5~T and 7~T are presented 
in Fig~\ref{fig_cp} (a).
The $C_p(T)$ of the non-magnetic analogue
compound \mto\ is plotted in (a) using red open circles.
A sharp $\lambda$-like transition at $T_N$ = 6.2~K is visible
for \cto.
External magnetic fields up to 7~T have negligible effect 
on the peak at $T_N$, however, an enhancement of $C_p(T)$ at 
$T < T_N$ is noticeable in the presence of externally applied
magnetic fields.
This may be related to the weak enhancement of 
ferromagnetism observed previously 
in $x$ = 0.7 \cite{baral2019low}.
As the Co magnetic lattice is diluted with Mg for 
$x\geq$0.1, the sharp feature at $T_N$ in \cto\ is replaced
by a broad feature in specific heat.
Figure~\ref{fig_cp} (b) shows the plot
of $C_p(T)/T$ versus temperature of all \cmto\ compositions
studied in the present work.
It is clear that with increasing Mg content ($x$), the
magnetic transition at $T_N$ is broadened. 
By taking the derivative, ${\mathrm d}C_p/{\mathrm d}T$, we determined
the transition temperatures for the $x$ = 0.1 and 0.3 compounds. 
For the $x$ = 0.1, anomalies in ${\mathrm d}C_p/{\mathrm d}T$ are present 
at 4.9~K and 3.2~K (see the inset of (a)), whereas for 
the $x$ = 0.3 case, at 4.5~K. 
Magnetization results from our previous work 
on the same composition had
shown multiple magnetic anomalies below 28~K in low applied 
fields which, upon application of external magnetic field,
smoothed into a broad transition \cite{baral2019low}.
The specific heat anomalies (in ${\mathrm d}C_p/{\mathrm d}T$) 
seen in the $x$ = 0.1 case matches well with our
previous magnetization results.
In the inset of (b), the $C_p(T)$ of the $x$ = 0.5 
compound in 0~T and 7~T (red curve) are plotted together. 
It can be seen that with the application of
magnetic field, there is a slight enhancement 
in the specific heat below $T_N$.  
The current specific heat data point towards the 
presence of ferromagnetic clusters probably of 
short-range ordered regions that get polarized
under the application of an external magnetic field.
\\
\indent
In order to obtain the magnetic contribution,
$C_m(T)$ towards total specific heat, we subtracted the 
$C_p(T)$ of \mto\ from that of \cto. 
Thus obtained $C_m(T)$ of \cto\ is presented in Fig~\ref{fig_cp2} (a).
Earlier studies on \cto\ have treated the Co atoms 
as a square net of Ising spins in order to model the
experimental specific heat \cite{kremer_specific_1988}.
We find that the model of a square net of
spins holds good in the present case as can be seen
by the solid line in (a), which is the specific heat of
an Ising net of spins.
Further, a broad feature is seen at $T>T_N$
reminiscent of a Schottky-type anomaly.
An attempt to model it using a 2-level Schottky term
leads to an acceptable fit as can be seen by the
blue solid line in (a).
Short-range spin correlations that exist above the
transition temperature giving rise to a broad
peak in specific heat is reported in CuSb$_2$O$_6$
for example \cite{nakua_crystal_1991, rebello_transition_2013}.
The characteristic energy scale determined from the
Schottky fit in the present case corresponds to $\Delta \approx$ 260~K.
Another plausible scenario that can support the 
Schottky-like feature in the specific heat is the
orbital magnetism that might be active in \cto \cite{christian_magnetic_2018}.
We further obtain support that the Schottky-like
peak in specific heat is closely linked to the
theoretical specific heat of $S$ = 3$/$2 1D-Ising model.
The calculations of specific heat for an $S$ = 3$/$2 
1D-Ising model predicts a peak at 1.25$R$, or about 10.4~J/mol-K and
should occur at $\mathrm{k_B}/J$ = 2.5 \cite{obakata1968one}. 
This closely matches the value at the top of the 
100~K-peak of $C_m(T)$ in Fig~\ref{fig_cp2} (a). 
Further, in the present case, $J/\mathrm{k_B}$ = $\textendash$40~K,
which might be linked to the interchain exchange constant.
Similar Schottky-like features are observed in the $C_m(T)$ of 
$x$ = 0.1$\textendash$0.7 compositions.
This supports the presence of short-range spin correlations
in all of the \cmto\ series.
Progressive Mg-doping disrupts the Co chains and leads to the
formation of clusters of spins displaying such Schottky-like
features in specific heat.
%
\begin{table}[!t]
	\caption{\label{tab_cp} 
		The Sommerfeld coefficient and Debye temperature of \cmto\ compounds 
		extracted from the analysis of low temperature specific heat.}
	\setlength{\tabcolsep}{12pt}
	\begin{tabular}{l l l l l l} \hline\hline
		$x$  & $\gamma$ (mJ/mol K$^2$)  & $\beta$ (mJ/mol K$^4$)   & $\theta_D$ (K) \\ \hline
		0  & 80.4(1)   & 0.0241(5)   & 892 \\
		0.1  & 55.4(3)   & 0.0254(3)   & 876 \\
		0.3  & 28.7(3)   & 0.0299(5)   & 829 \\
		0.5  & 13.3(7)   & 0.0191(7)   & 963 \\
		0.7  & 4.4(1)   & 0.0211(5)   & 931 \\ \hline\hline 
	\end{tabular}
\end{table}
The $C_m(T)$ of the other compositions in the \cmto\ series are
shown in panel (b) of the figure. 
Evidently, the magnetic contribution
to the specific heat decreases with increasing Mg content.
The magnetic entropy that is obtained using
$S_m$ = $\int_{0}^{T}\! {\mathrm d}C_m/T{\mathrm d}T$ is plotted for all 
compositions in (c). 
Only 6$\%$ of the total spin-only entropy 
of $R$ln(2S + 1) ($S$ = 3/2 and $R$ is the universal gas constant) 
is released close to the $T_N$.
The low magnetic entropy release observed close to 
the magnetic ordering temperature indicates that significant 
amount of short-range magnetic order
exist above the $T_N$ in \cto\ and the Mg-doped
compounds \cite{christian_local_2015}.
\\
\indent
Further, we have analyzed the low temperature specific heat 
of \cmto\ using the expression, $C_m(T)$ = $\gamma T$ + $\beta T^3$. 
The fit parameters and the values of Debye temperature estimated for 
the different compositions are given in Table~\ref{tab_cp}. 
The Debye temperature was estimated
using the equation, $\theta_D$ = $\left(12p\pi^4R/5\beta\right)$, 
where $p$ is the number of atoms in the unit cell and $R$ is 
the universal gas constant.
The values of $\gamma$ shows a gradual decrease as a 
function of composition, $x$; whereas the $\beta$ 
values show an anomaly for $x$ = 0.5. 
The value of $\gamma$ that is observed in \cto\ lies intermediate
with the values of 203~mJ/mol~K$^2$ for CoSb$_2$O$_6$ and 58.7~mJ/mol~K$^2$
for CuSb$_2$O$_6$ \cite{christian_local_2015}. 
The insulating nature of these compounds might point 
to the fact that the $\gamma$ term signifies the 
entropy associated with 
short-range order
The magnetic specific heat in the temperature 
range below $T_N$ could be modeled using
\begin{equation}
C_m(T) = \gamma_\mathrm{AFM} T + A_2~\mathrm{exp}(-\Delta/k_\mathrm{B}T).
\label{eqn1}
\end{equation}
The first term on the right-hand-side of this expression
is the specific heat of free electrons near the Fermi level and
the second term originates from the electrons that participate
in the magnon dispersion.
A fit using the above expression to the experimental 
magnetic specific heat below $T_N$ leads
to the values $\gamma$ = 0.018(3), $A_2$ = 0.06(5) and
$\Delta/k\mathrm{B}$ = 20.2(5)~K.
This fit is presented in the inset of Fig~\ref{fig_cp2} (a)
as a thick solid line.
\subsection{Raman scattering and thermal conductivity}
\begin{figure}[!b]
	\centering
	\includegraphics[scale=0.38]{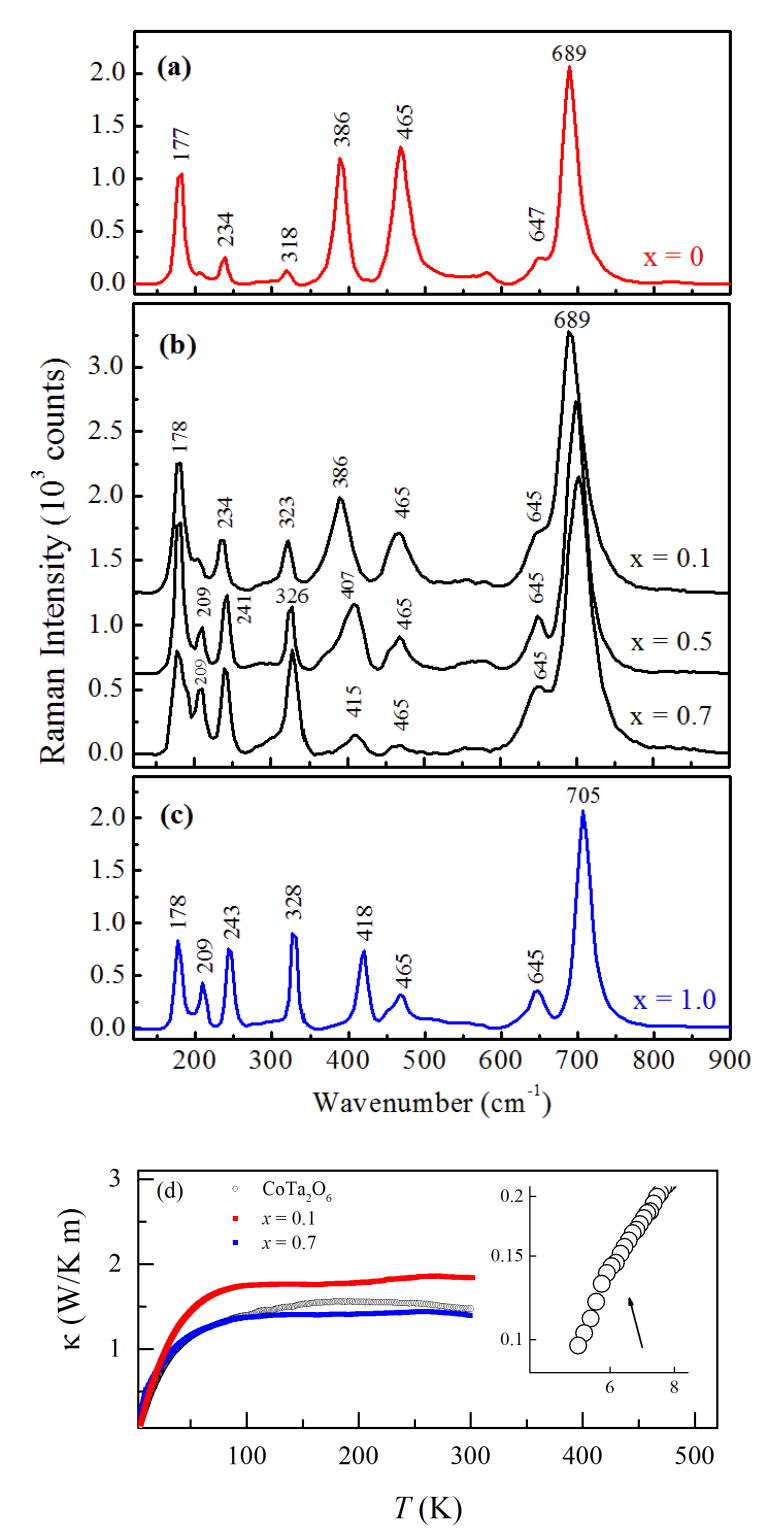}
	\caption{ \label{fig:kappa} (color online) 
		Raman scattering intensity
		of \cmto\ for (a) $x$ = 0, (b) $x$ = 0.1, 
		0.5, 0.7 and (c) $x$ = 1.0
		at $T$ = 300~K. 
		The thermal conductivity, $\kappa (T)$ of the three 
		compositions of \cmto: $x$ = 0, 0.1 and 0.7.
		The phase transition at $T_N \approx$ 6~K in the case of $x$ = 0
		is shown enlarged in the inset. }
\end{figure}

For a better understanding of the Mg-doping 
effects on the structural features of \cto, 
we present in Fig~\ref{fig:kappa} (a-c) the 
room-temperature micro-Raman spectra of \cmto\ 
with x = 0, 0.1, 0.5, 0.7, and 1.0.  
Although similar Raman vibrational lines for \cto\ and \mto\ 
($x$ = 0.0 and 1.0) have been reported in the literature
\cite{husson1979spectres, haeuseler1981infrared}, 
some repetition of results in research contributes 
to the supportive purposes of cross-checking and 
confirmation of reproducibility.
Since the Raman spectra of tantalates are dominated 
by modes corresponding to Ta$_2$O$_6$ units, 
with the bivalent metal atoms making a much smaller 
contribution, a comparison between the Raman 
results presented in Fig~\ref{fig:kappa} (a) and 
(c), indeed, reveals an expected similarity of vibrations 
\cite{husson1979spectres}. 
However, definite trends are observed in 
Fig~\ref{fig:kappa} (b) with Mg atoms progressively 
substituting for 
Co atoms and resulting in elongations of Co$\textendash $O 
bonds and angle modifications of the cyclic 
Ta$\textendash$O$\textendash \mathrm{Co}\textendash$O$\textendash$Ta 
structure (as Ta$\textendash$O bonds will couple differently 
with Co$\textendash$O bonds after substitution). 
Direct evidence of Mg incorporation is the 
increase in the intensity of the 209~cm$^{-1}$ Raman line, 
which corresponds to vibrational modes of 
Mg$\textendash$O bonds 
(see the Raman spectrum of \mto\ in Fig~\ref{fig:kappa} (c)). 
Furthermore, frequency shifts and intensity 
changes are also seen in all \cmto\ Raman 
spectra for other vibrational lines. 
For example, with Mg-doping, the Raman peak 
at 234~cm$^{-1}$ in the spectrum of \cto\ (see Fig~\ref{fig:kappa} (a)) 
increases in intensity and shifts to a higher 
frequency, to a final value of 242~cm$^{-1}$ 
in the Raman spectrum of \xseven, 
a value that is very close to the corresponding 
one seen in the Raman spectrum of \mto. 
The Raman feature at 318~cm$^{-1}$ in the spectrum 
of \cto\ shifts to 328~cm$^{-1}$ in the spectrum of \xseven. 
It also exhibits a similar intensity increase. 
The 386~cm$^{-1}$ Raman line, too, presents higher frequency 
shifts, to 415~cm$^{-1}$ in the Raman spectrum of 
\xseven; it also decreases in intensity.
A similar decrease in intensity is observed for the 
Raman peak at 465~cm$^{-1}$ 
without a  frequency shift. 
Comparison of the intensity of this 
vibrational line in the Raman spectrum of \cto\ 
with its intensity in the spectrum of 
\mto\ reveals that the observed pattern 
in intensity changes in the 
\cmto\ Raman spectra again correlates with 
the amount of Mg-doping. 
The Raman vibrational lines at 
234 and 318~cm$^{-1}$ in the \cto\ 
spectrum have lower intensities than those 
at 243 and 328~cm$^{-1}$ 
in the \mto\ spectrum. 
On the other hand, the Raman bands at 
386 and 465~cm$^{-1}$ in the \cto\ spectrum have 
higher intensities than those at 418 and 
465~cm$^{-1}$ in the \mto\ spectrum. 
Thus, with more Mg incorporation, 
the Raman spectra of  \cmto\ are expected to 
resemble more that of \mto\ and less that of
\cto. 
Finally, the frequency at 689~cm$^{-1}$ 
associated with Ta$\textendash$O bond vibration in the 
\cto\ Raman spectrum (Fig~\ref{fig:kappa} (a)) 
shifts slightly in the \cmto\ spectra, to a value of 702~cm$^{-1}$ 
for the case with the highest amount of Mg-doping. 
Thus, the Raman results obviously demonstrate 
morphological changes of the \cto\ structure with Mg-doping.
\\
\indent
The temperature dependence of thermal conductivity, 
$\kappa (T)$, measured on the polycrystalline
samples of \cto, \xone\ and \xseven\ is shown in the main panel of
Fig~\ref{fig:kappa} (d).
In the inset of the figure, the low temperature part of 
$\kappa (T)$ for \cto\ in the vicinity of the magnetic
phase transition $T_N$ = 6.2~K is shown.
The value for $T_N$ observed in $\kappa (T)$ is similar
to the value observed in $C_p (T)$.
Recent detailed work on the thermal conductivity of 
\cto\ and CoSb$_2$O$_6$ single crystals showed that the Ta
compounds have lower values of $\kappa(T)$ compared to the
Sb counterparts \cite{prasai2018resonant}.
It is worth noting that the thermal conductivity of our \cto\ sample
could be reduced due to the porosity that might be present in the material.
Furthermore, a sharp upturn of $\kappa(T)$ is observed in \cto\ 
at the Ne\'{e}l temperature. 
In our polycrystalline sample, this transition is not so sharp 
as observed in the single crystals but is clearly seen as a 
kink in the $\kappa(T)$ curve. 
The thermal conductivity features observed in the present case
of \cto\ confirm the fact that the strong phonon scattering 
is coupled to the short-range antiferromagnetic order that
develops at temperatures above the Ne\'{e}l temperature.
$\kappa(T)$ is very low and nearly temperature independent
above 100~K.
The low value of $\kappa(T)$ probably indicates 
low charge carrier density and the temperature independent
nature reflects very efficient scattering of heat carrying
phonons, which might be comparable to phonon-glass systems.  
Detailed thermal transport studies on single crystals might be helpful
to conclude on this topic.
\subsection{Neutron diffraction}
\begin{figure}[!b]
	\centering\includegraphics[scale=0.20]{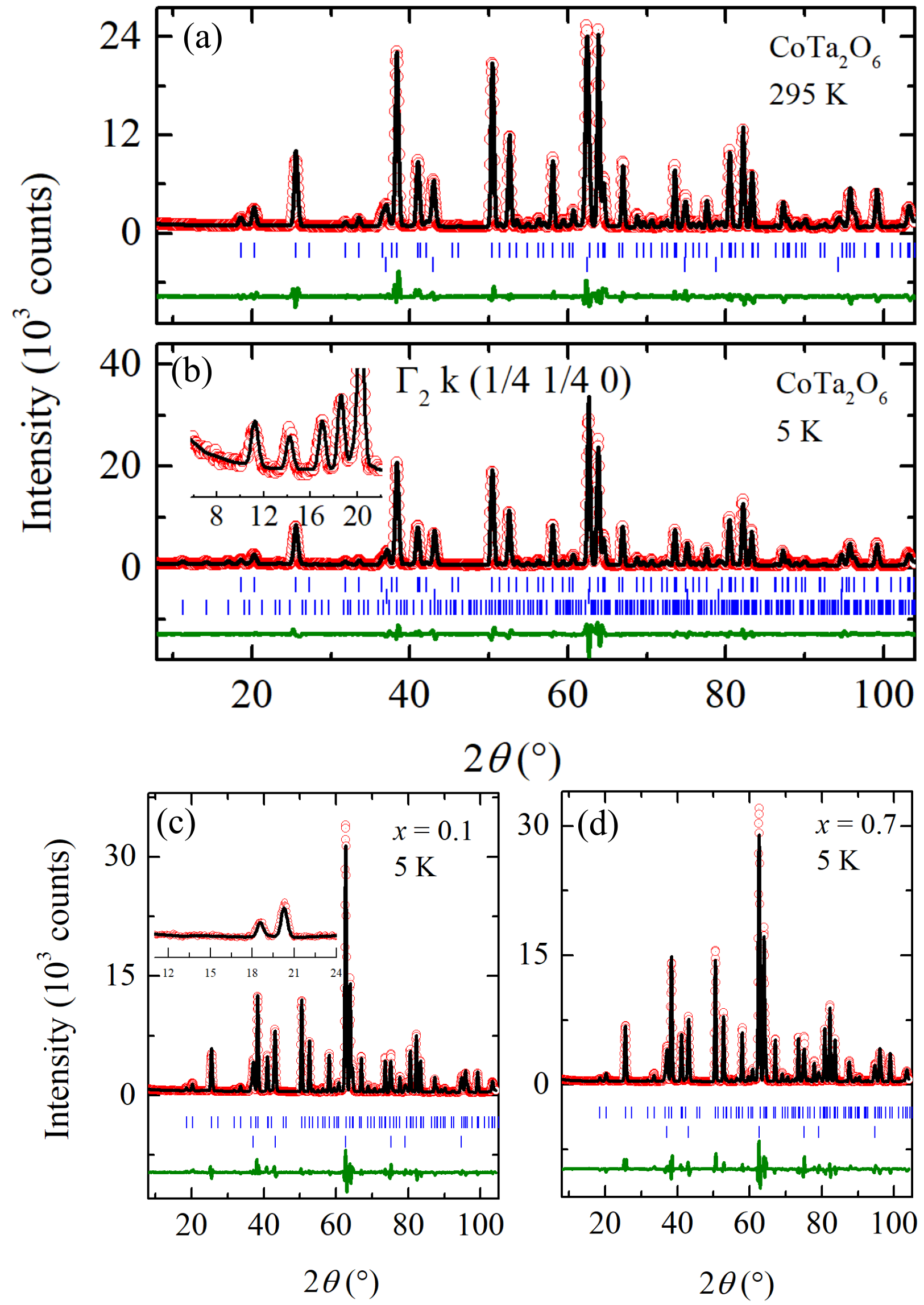}
		\centering\includegraphics[scale=0.08]{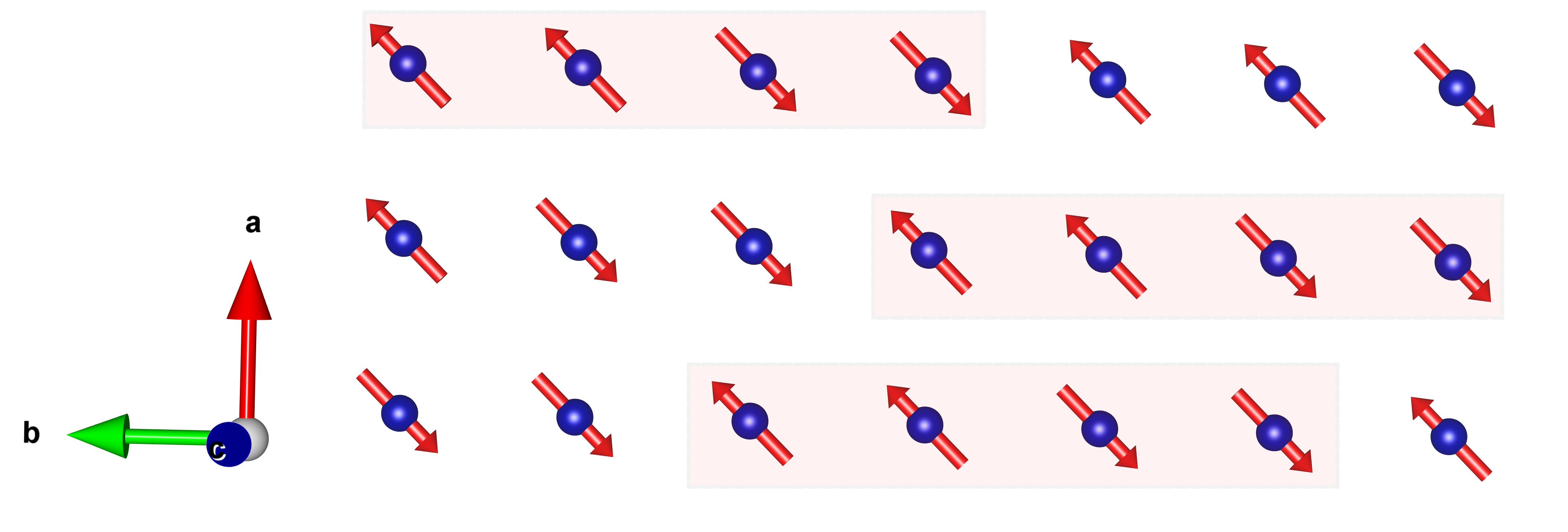}
	\caption{(color online) \label{fig:npd1} 
		Top: (a) The neutron powder diffraction pattern of \cto\ at 
		295~K refined in the $P4_2/mnm$ space group. 
		(b) Shows the pattern at 5~K where the magnetic peaks 
		are shown magnified in the inset. 
		$\Gamma_2$ representation with a propagation vector 
		$\left(\frac{1}{4} \frac{1}{4} 0\right)$ accounts for the magnetic peaks.
		Nuclear Bragg peaks from the Al sample holder 
		used for the experiment are accounted for in all cases. 
		(c, d) Shows the patterns for $x$ = 0.1 and 0.7, 
		respectively, at 5~K.  
Bottom: A representation of a layer of Co magnetic 
moments in the $ab$ plane showing the
$\left(+ + \textendash~\textendash \right)$ arrangement.}
\end{figure}

\indent 
The neutron powder diffraction patterns of \cmto\ were recorded
at the PSD instrument at different temperatures between 295~K and 5~K.
Figure~\ref{fig:npd1} (a, b) show the neutron powder 
diffraction patterns obtained for the $x$ = 0 
compound ($i.e.,$ \cto) at 295~K and 5~K, respectively. 
The crystal structure of \cto\ is well-documented 
in the literature \cite{antonietti_structure_2001}
as trirutile type of crystal structure. 
All the compositions of \cmto\ in the present study were
identified to retain the trirutile space group $P4_2/mnm$
at all temperatures. 
In Fig~\ref{fig:npd1}, the red circles represent the experimentally 
recorded pattern while the black solid line is the model 
fit using $P4_2/mnm$ space group. 
In the trirutile structure of \cto, the Co$^{2+}$ and Ta$^{5+}$ 
cations are surrounded by O$^{2-}$ octahedra, 
and successive Co–O planes (at $z$ = 0 and $z$ = 1/2) 
are separated by two Ta$\textendash$O planes. 
The refined Co$\textendash$O bond distances 
and the O$\textendash$Co$\textendash$O 
bond angles are given in the Table~\ref{tab:str}
along with the refined atomic parameters and lattice
constants. 
The refined lattice parameter at $T$ = 295~K 
is comparable to the value reported earlier in a
structural report\cite{antonietti_structure_2001}. 
We did not observe any structural phase 
transformation in \cto\ as a function of 
temperature down to 5~K or as a function of
composition, $x$, in the case of \cmto. 
The $x$-dependence of both $a$ and $c$ showed 
a weak anomaly at $x$ = 0.5 (not shown) which was reflected
in the bond parameters as well.
The most recent report on the neutron diffraction
study of \cto\ reports on a polycrystalline sample
with small impurities, CoO and Ta$_2$O$_5$ \cite{kinast_magnetic_2010}.
Though the crystal structure is refined in the 
$P4_2/mnm$ space group, the lattice constants reported
($a$ = 4.715(9)~{\AA}, $c$ = 9.127(6)~{\AA} at 1.5~K)
are different from the values obtained in the present work.
\\
\begin{table}
	\caption{\label{tab:str} 
		The structural parameters of \cto\ at 295~K and 5~K
		obtained from the refinement of the neutron powder diffraction data
		are presented in the top two rows. 
		The refined lattice parameters at $T$ = 295~K are $a$ ({\AA}) = 4.7382(9), 
		$c$ ({\AA}) = 9.1758(2) and at $T$ = 5~K are $a$ ({\AA}) = 4.7421(3) 
		and $c$ ({\AA}) = 9.1791(3). 
		The bond parameters of \cmto\ at 295~K as a function of $x$ are presented in the
		last row of the table.}
	\setlength{\tabcolsep}{15pt}
	\begin{tabular}{l l l l l l l} \hline\hline
		$T$ = 295~K  & $x ({\AA})$  & $y ({\AA})$  & $z ({\AA})$ \\ \hline
		Co ($2a$) & 0 & 0 & 0 \\
		Ta ($4e$) & 0 & 0 & 0 \\
		O ($4f$) & 0.3118 & 0.3118 & 0 \\
		O ($8j$) & 0.2960 & 0.2960 & 0.3230 \\ \hline \hline
		$T$ = 5~K  & $x ({\AA})$  & $y ({\AA})$  & $z ({\AA})$ \\ \hline
		Co ($2a$) & 0 & 0 & 0 \\
		Ta ($4e$) & 0 & 0 & 0 \\
		O ($4f$) & 0.3115 & 0.3115 & 0 \\
		O ($8j$) & 0.2957 & 0.2957 & 0.3236 \\ \hline\hline 
	\end{tabular}
	\setlength{\tabcolsep}{10pt}
	\begin{tabular}{l l l l l l}
		$x$  & Co$\textendash$O1 ({\AA})  &  Co$\textendash$O2 ({\AA}) & O2$\textendash$Co$\textendash$O2 ($^\circ$) \\ \hline
		0 & 2.0815(19) & 2.1164(16) & 80.52(8) \\
		0.1 & 2.088(3) & 2.1156(17) & 80.58(9) \\
		0.3 & 2.0872(19) & 2.1169(16) & 80.39(8) \\
		0.5 & 2.084(3) & 2.1104(17) & 80.08(9)  \\
		0.7 & 2.078(3) & 2.113(3) & 79.84(12)  \\ \hline\hline 
	\end{tabular}
\end{table}
\indent
With the reduction in temperature to 5~K, 
additional Bragg peaks at 2$\theta \approx$ 11.3$^\circ$,
14.2$^\circ$, 11.3$^\circ$, 17$^\circ$ and 19$^\circ$
are observed in the neutron powder diffraction pattern.
The additional peaks, that correspond to the magnetic order
developing in the compound below the $T_N$, are shown 
enlarged in the inset of Fig~\ref{fig:npd1} (b). 
In order to solve the magnetic structure, a profile fit and
$\mathrm k$-search was performed using the 
utilities within Fullprof Suite software.
A $k$-value of $\left(\frac{1}{4}, \frac{1}{4}, 0\right)$ 
was obtained.
Using the representation analysis tool in 
SARA$h$, we then obtained the symmetry-allowed magnetic 
representations for the space group $P4_2/mnm$. 
Accordingly, the magnetic structure was determined in
$\Gamma_2$ representation. 
The Co magnetic moments adopt a magnetic arrangement 
$\left(+ + ~\textendash ~\textendash \right)$ 
that constitutes a sinewave modulated structure. 
The average magnetic moment on the Co sites is obtained as
2.3(2)~\muB. 
The magnetic structure presented here is similar to that 
put forward in the case of weakly diluted
Fe$_{1-x}$Co$_x$Ta$_2$O$_6$ \cite{kinast_bicriticality_2003}.
The magnetic structure of FeTa$_2$O$_6$ is described
in terms of three dimensional stacking of
antiferromagnetic planes, where the
anisotropy of one set of planes is rotated 90$^\circ$
with respect to the other \cite{eicher1986magnetic}.
On the other hand, the magnetic structure of
\cto\ is close to that of a complex helix with
components on the base plane and along the
$c$ axis \cite{reimers_crystal_1989}.
Hence a solid solution of \cto\ and FeTa$_2$O$_6$
resulted in a antiferromagnetic structure
with competing interactions, where, intermediate
compositions possess two different magnetic propagation
vectors.
The magnetic structure solution presented here
can be connected to the specific heat
of \cto.
The $\left(\frac{1}{4}, \frac{1}{4}, 0\right)$ propagation vector 
leads to a situation where neutron data are not able
to differentiate between a sequence $\left(+ + \textendash~\textendash \right)$,
where all magnetic moments are equal, or a situation $\left(+~0~\textendash~0\right)$ 
where two out of four spins have a large moment and the other two, 
no magnetic moment. 
Hence, the magnetic structure estimated here aligns well
with the observation in an earlier work where the
specific heat below $T_N$ suggested that a large
fraction of spins remained disordered \cite{christian_magnetic_2018}.
\\
\indent 
The neutron powder diffraction patterns obtained for $x$ = 0.1 and 0.7
at 5~K are presented in Fig~\ref{fig:npd1} (c) and (d) respectively.
The low-angle region of the pattern is shown
magnified in (c) to clearly show that the magnetic Bragg peaks 
are absent even at Mg-dilution of 10$\%$. 
The temperature dependence of magnetization of $x$ = 0.1 composition
revealed anomalies at low temperatures \cite{baral2019low}.
The present results obtained through neutron diffraction
confirms that the magnetism in \cto\ is completely suppressed
upon increasing the concentration of Mg to 10$\%$.
The neutron diffraction patterns of $x$ = 0.3, 
0.5 and 0.7 at 295~K and 5~K are similar to that of $x$ = 0.1
confirming the absence of long-range magnetic ordering
in \cmto\ for $x > 0.1$.
The current results align well with the
bulk magnetization data \cite{baral2019low}
except for an enhancement of ferromagnetic correlations
found in the heavily doped composition ($x$ = 0.7), 
however, in external magnetic fields.
We do not observe experimental indications of 
diffuse magnetic order in any of the \cmto\ compositions.
\subsection{Density functional theory}
We first consider the geometrical optimization for the 
experimental parameters obtained from our diffraction 
experiments as tabulated in Table~\ref{tab:str} for $T$ = 5 K.
\begin{figure*}[!t]
	\centering
	\includegraphics[scale=0.18]{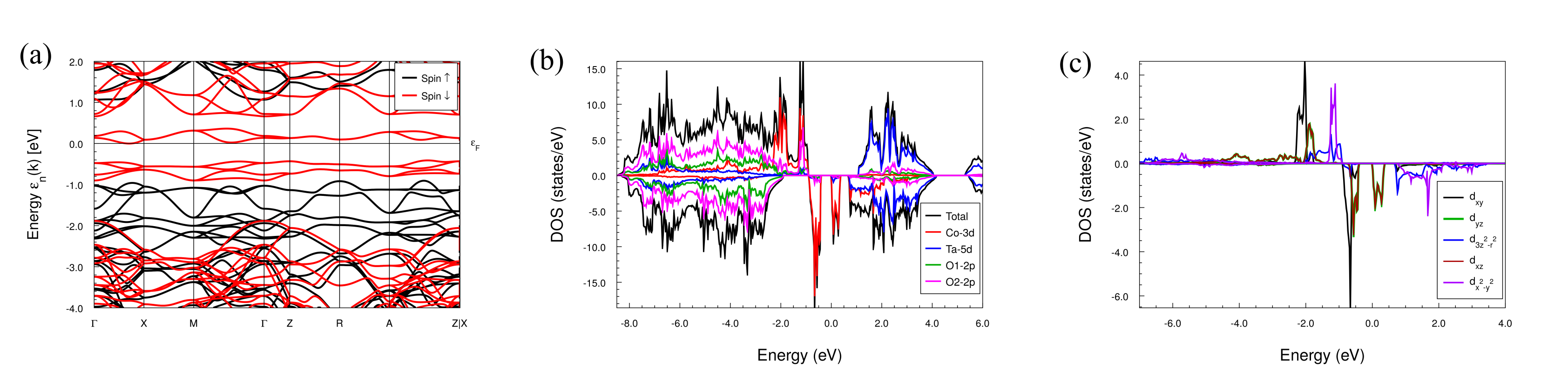}
	\caption{ \label{fig_dft} (color online) 
		(a) Scalar-relativistic band structure within GGA for the 
		FM state of \cto. Bands in black (red) indicates the spin-up (spin-down) 
		channels. 
		(b) Total (black) and partial DOS for Co (red), Ta (blue), 
		apical-O (green) and planar-O (pink) of \cto~within GGA.
		(c) Orbital resolved DOS for Co-3d states in \cto.}
\end{figure*}
All the results presented here are from the relaxed structure. 
The deviation of the obtained data are negligible 
(i.e., $\leq2\%$ change in the experimental positions) 
compared with the experimental lattice parameters.
We first consider the electronic properties of a
non-magnetic configuration by DFT calculations, 
from where we observe a sharp peak at the Fermi level 
($E_{\rm F}$) indicating the magnetic instability of \cto. 
Based on this, we further investigate the magnetic ground 
state of \cto, using the optimized parameters as mentioned above. 
The total energy calculations for the collinear 
ferromagnetic (FM-$\uparrow\uparrow$), 
antiferromagnetic (AFM-$\uparrow\downarrow$), 
and non-magnetic (NM) configurations were calculated. 
The ground state was found to be FM where the energy 
difference between the lowest total energy state 
and the first excited AFM state amounts to just 
5.77~meV per unit cell. 
The small energy difference between FM and AFM 
clearly indicates the competing ground state 
and the reasonable explanation to the AFM 
ground state below 5~K as observed here 
in the experiment. 
The FM ground state obtained
through the DFT calculations is obtained
by using the structural parameters
reported by some of us recently \cite{baral2019low}.
\\
\indent 
We further consider the magnetic anisotropy energy 
(MAE) of \cto\ for the FM ground state. 
The easy axis was found along the (100) direction 
with MAE of approximately 2.5~meV per unit cell. 
In the calculations, each Co couples ferromagnetically 
with its neighbor Co via oxygen and Ta in the 
trirutile structure. 
In \cto, each Co with charge state +2 formally has 
a 3$d^7$ configuration. 
This is expected to  carry a magnetic moment 
of $\pm$3 \muB~ within an ionic picture, 
while our first-principles calculation 
indicates the spin moment of 2.68~\muB\ and an 
orbital moment of 0.17~\muB\ per Co. 
Although finite strength of spin-orbit coupling 
(SOC) has been observed on the Co-site, it’s 
influence at and around $E_{\rm F}$ 
in the electronic structure is negligible. 
Because of the charge transfer effects between 
Co(Ta) and O atoms, Ta and oxygen gain a partial 
moment of 0.048~\muB\ per Ta, 0.032 \muB\ per planar 
oxygen and 0.041~\muB\ per apical oxygen atoms. 
The difference between two in-equivalent atoms 
are mainly due to their different bond lengths 
(Co$\textendash$O bond length: apical (planar) 2.053 (2.093)~\AA) 
arranged with the Co atoms. 
The obtained moment of Co is found to be in good 
agreement with our experimental moment of 2.13~\muB/Co. 
The small difference is attributed to the environmental 
effects (temperature, etc.), SOC and the hybridization 
among the Ta and oxygen atoms.\\
\indent
We performed the DFT calculations 
using the crystal parameters obtained at 11~K. 
This data suggests the FM ground state.
Within the GGA calculations, \cto\ is found 
to be an indirect band gap semiconductor as 
shown in Fig~\ref{fig_dft} (a). 
The band gap found in spin-up channel is $~2.0$~eV, 
while that in spin-down channel it is $\sim0.47$~eV 
respectively. 
As observed from the total and partial DOS in 
Fig~\ref{fig_dft} (b), the dominant contributions 
to the total DOS in spin-up channel at and below $E_{\rm F}$ 
(between $\textendash$2.5~eV to 0.0~eV) are mainly 
from the Co-3$d$ orbitals hybridizing with the O$\textendash$2$p$ 
states (see Fig~\ref{fig_dft} (b)) while above $E_{\rm F}$ ($<$1.5~eV) 
are the contributions from Ta 5$d$ states. 
This clearly shows the full occupancy of Co 5$d$ states 
in spin-up channel (see also the spin-up band structure 
below $E_{\rm F}$ where five $d$-bands are clearly 
distinct with two degenerate bands overlapping at 
high symmetry points such as X, M, etc. below $\textendash$1~eV). 
In the spin-down channel on the other hand, 
major contributions below $\textendash$1.5~eV are from the O 2$p$ 
states while two occupied $d$ state from Co 3$d$ 
orbitals are found with sharp peaks between 
$\textendash$1 and $\textendash$0.5~eV hybridizing with the O 2$p$ states;
while an isolated $d$ band from the $t_{2g}$ orbitals
lie just above the $E_{\rm F}$. 
Due to crystal field effect, the two empty bands from Co $d(e_g)$ 
lie around 1.3~eV above $E_{\rm F}$ hybridizing with the Ta-5$d$ orbitals.
These features can be observed distinctly from the band 
structure shown in Fig~\ref{fig_dft} (a) and the local-orbitals 
from Co 3$d$ states (see Fig~\ref{fig_dft} (c)).  
Since Co$^{2+}$ has five valence electrons occupied 
in the spin-up and only two $t_{2g}$ states  is expected 
to be occupied in spin down channel, the 
remaining three empty orbitals should lie 
above the conduction band. 
The partial DOS result shown in Fig~\ref{fig_dft} is 
found consistent with the idea that the 
Co $d$($t_{2g}^3e_g^2$) spin-up channel is fully 
occupied, lying below -1~eV from the $E_{\rm F}$. 
In spin down channel, two of the $t_{2g}$ are 
occupied close to $E_{\rm F}$ while one empty $t_{2g}$ 
and two empty $e_g$ states appear 
above $E_{\rm F}$ in the conduction region. 
The corresponding band structure is shown in Fig~\ref{fig_dft} (b). 
As observed, in the scalar relativistic mode (absence of SOC), 
\cto\ has a band gap of 2.0~eV in the spin-up channel, 
while that in the spin-down channel is $\sim$0.47~eV. 
In contrast, when SOC is considered, the gap-size is 
reduced to $\sim$0.41~eV (not shown). 
The noticeable change is the shifting of 
$E_{\rm F}$ from the bottom of the conduction band to the 
top of the valence band. 
\\
The specific heat data presented in this work point towards disordered
ferromagnetic spin clusters in \cto.
The neutron diffraction analysis supports the above
scenario where the $\left(+ + \textendash~\textendash \right)$ or
$\left(+~0~\textendash~0\right)$ structures are indicated.
Slight deviation from nominal stoichiometry due to volatility of Co might drive 
certain degree of disorder in \cto.
This may, in turn, influence stabilization of different valences for Co which need to be
probed in detail using high resolution X-ray spectroscopy.
The vacancies of Co that may have occurred during the synthesis process
can indirectly influence oxygen stoichiometry of \cto.
In such a scenario, we can expect to observe FM similar to
carbon-doped ZnO or aluminum-doped TiO$_2$ thin films \cite{referee1, referee2}. 
Ferromagnetism is expected to be highly favorable in the present case as \cto\ has already shown 
magnetism in the absence of vacancies, as per our DFT results.
Oxygen vacancy at particular site can be expected to transfer hole either to Ta or Co-site. 
The vacancy can thus be expected to modify the electronic as well as the magnetic properties of 
\cto\ however, a direct correlation of such results do not exist as of now.
Single crystals of another trirutile, MgTa$_2$O$_6$, prepared using the 
technique of optical floating zone under argon was reported as appearing black due to
oxygen vacancies \cite{higuchi1993growthMgTa2O6}. 
However, synthesis in air helps in achieving a yellow color for the crystal with less
oxygen vacancies \cite{xu2019ramanMgTa2O6}.
The intrinsic atomic defects due to oxygen off-stoichiometry in trirutiles
has been recently treated using detailed simulations \cite{tealdi2004defectMgTa2O6}.
The oxygen positions in the trirutile structure has two
Wyckoff positions and that the most favourable
energy is associated with the vacancies on the  O1 site.
However, the calculated defect energies suggested that they
are insignificant in the case of MgTa$_2$O$_6$.
In \cto, similarly, we do not anticipate significant changes in the magnetic 
properties due to slight alterations from nominal stoichiometry.
\section{Conclusions}
Signatures of low dimensionality and short-range
magnetic order are observed in the trirutile series,
\cmto.
The specific heat of \cto\ reveals a magnetic phase
transition at 6.2~K and a broad feature at $T>T_N$, ascribed
to short-range 
magnetic order.
The $T_N$ is relatively robust in applied magnetic fields but
is reduced upon Mg-doping at the Co site.
However, the broad feature related to short-range spin
order is present in all of the doped compositions.
A low value of entropy is released at $T_N$.
Analysis of specific heat support low dimensional
features of magnetism and short-range order in \cmto\ trirutiles.
Mg-doping of 10$\%$ is sufficient to destabilize the
antiferromagnetic order seen in \cto.
Performing neutron diffraction on a phase-pure sample, we
have determined the magnetic structure of \cto\ to be
a sine wave-type order with propagation vector 
$k$ $\left( \frac{1}{4}, \frac{1}{4}, 0\right)$.
Density functional theory studies predict the ferromagnetic 
ground state, however, the negligible (small) total energy 
differences between the ferromagnetic and antiferromagnetic 
configuration indicates a competing ground magnetic state.
This in turn supports the short-range order we observe in
specific heat data.
Our computational results supports the observation of 
strong anisotropy effects in the class of trirutile compounds.
%
\section{Acknowledgements}
HSN acknowledges the UTEP start-ups and the Rising-STAR award from UT system
in supporting this work. 
KG acknowledges support from the DOE’s Early Career Research 
Program. 
NP acknowledges support from INL's LDRD program (18P37-008FP). 
MPG acknowledges the Alexander von 
Humboldt Foundation, Germany for the financial support, 
and Higher Education Reform Project of 
Tribhuvan University for the start-up grant. 
MPG also thanks Ulrike Nitzsche for technical assistance. 
BPB thanks Sainamaina-Municipality, Nepal for the 
partial support, and SB thanks UGC-Nepal for the 
scholarship support under MRS-2075/76 program. 
AMS thanks the SA NRF (93549) and the URC/FRC of UJ for financial assistance.
\\
\\
$^\dagger$ Contributed equally\\

 \end{document}